\documentclass{webofc}
\usepackage{graphicx,color}
\usepackage[varg]{txfonts}   
\begin{document}
\title{Field induced phase transition in the few photon regime}
%
%

\author{
\lastname{A.~D.~Panferov}\inst{1}\fnsep\thanks{\email{panferovad@info.sgu.ru}} \and
       \lastname{S.~A.~Smolyansky}\inst{1}
             \and
        \lastname{A.~I.~Titov}\inst{2}
             \and
        \lastname{B.~K\"ampfer}\inst{3,4}
             \and
	 \lastname{A.~Otto}\inst{3,4}
             \and
	 \lastname{D.~B.~Blaschke}\inst{2,5,6}
             \and
	 \lastname{L.~Juchnowski}\inst{5}
}

\institute{Saratov State University, Saratov, Russia \and
		Joint Institute for Nuclear Research,  Dubna, Russia \and
     	        Helmholtz-Zentrum Dresden-Rossendorf, Germany \and
		Institut f\"ur Theoretische Physik, Dresden, Germany\and
		University of Wroclaw,  Wroclaw, Poland \and
		National Research Nuclear University (MEPhI), Moscow, Russia
                     }

\abstract{
Some features of the field induced phase transition accompanied by the vacuum creation of an electron-positron plasma (EPP) in strong time-dependent electric fields have been discussed in the work \cite{1} 
in the domain of the tunneling mechanism ($\omega \ll m$, where $\omega$ is the characteristic frequency of the external field and $m$ is the electron mass).
In the present contribution the features of the this process will be considered in the few photon domain where $\omega \sim m$.
We observe a narrowing of the transient domain of the fast oscillations and, mainly, a considerable growth of the effectiveness of the EPP production.
Under these circumstances, we see an increase of the effectiveness of the EPP creation in the particular case of a bifrequent excitation, where both mechanisms (tunneling and few photon) act simultaneously \cite{2,3}.
}
\maketitle
\section{Introduction}
\label{intro}

The present note is devoted to the investigation of some features of the dynamical Schwinger effect which concerns the creation of an electron-positron plasma (EPP) from the vacuum in fastly varying spatially homogeneous external electric fields.
Initially, the Schwinger effect was predicted for a constant electric field \cite{4,5,6}.
According to these estimates a noticeable vacuum creation of electron-positron pairs in a strong electric field is possible only near the very high critical value for the field strength $E_c=m^2/e=1.3\times10^{16}$ V/cm.
For this reason, the interest in this problem is linked to a practical basis provided, e.g., by the further development of ultra-high intensity laser systems.
The characteristic frequency of the pulse  for a laser in the optical range is considerably smaller than the electron mass (using natural units $\hbar=c=1$).
A virtual pair must accumulate the energy of a very large number of photons of the external field before being "created".
The dependence of the efficiency of the pair production on the field strength has approximately the same character as in the case of a constant field.
Therefore the hope for detecting the creation of an EPP created from vacuum by the dynamical Schwinger effect can be realized only for field strengths in the immediate vicinity of $E_c$.

Encouraging predictions of enhanced efficiency of EPP vacuum excitation for the assisted Schwinger effect using the bifrequent external field of two lasers stimulate the interest to study the high frequency domain (few photon area) in the single-frequency regime.
Below we discuss some preliminary results of research on the field induced phase transition in this domain using the kinetic equation (KE) approach \cite{7} as the basis.
A brief description of this tool is given in Sect.~\ref{kinetic}.
Some features of the phase transition under the influence of a fastly alternating external field from an initial vacuum state to the out-state of residual EPP are presented in Sect.~\ref{transit}.
We put emphasis on comparison with the tunneling domain.
On the examples of the simplest models for pulsed fields we demonstrate a significant increase of the efficiency of the EPP creation in the few photon domain for both, the distribution function and the observable pair density in the final state.
The reason for this enhancement is the Breit-Wheeler mechanism.
A second feature is the significant narrowing of the transient domain of fast oscillations.
A discussion of the results is given in Sect.~\ref{discuss}.

\section{Kinetic equation}
\label{kinetic}

The creation of the EPP in a homogeneous, linearly polarized electric field $E(t)=-\dot{A}(t)$ with the vector potential (in Hamiltonian gauge) $A^\mu(t)=(0,0,0,A(t))$ is described by the KE \cite{7} for the distribution function 
\begin{equation}
\label{ke}
\dot f(\mathbf{p} ,t) = \frac{1}{2} \lambda(\mathbf{p},t )\int\limits^t_{t_0} dt^{\prime} \lambda(\mathbf{p}, t^{\prime}) [1-2f(\mathbf{p}, t^{\prime})]\cos\theta(t,t^{\prime}),
\end{equation}
where
\begin{equation}
 \lambda(\mathbf{p},t) = e E(t)\varepsilon_{\bot}/\varepsilon^{2}(\mathbf{p},t),  ~~~ 
 \theta(t,t^{\prime}) = 2 \int^t_{t^{\prime}} d\tau \, \varepsilon (\mathbf{p} ,\tau). \nonumber
 \end{equation}
Here $\lambda$ is the amplitude of the vacuum transitions, and $\theta$ is the high-frequency phase. 
The quasienergy $\varepsilon(\mathbf{p} ,\tau)$, the transverse energy $\varepsilon_\bot$ and the longitudinal quasimomentum $P(t)$ are defined as
\begin{equation}
\varepsilon(\mathbf{p} ,t) = \sqrt{\varepsilon^2_{\bot}(\mathbf{p}) + P^2(t)} ,
~~~
\varepsilon_\bot = \sqrt{m^2 + p^2_\bot},
~~~
 P(t) = p_\parallel -eA(t).
\nonumber
\end{equation}
Here $p_\bot$ and  $p_\parallel$ are the  momentum components perpendicular and parallel to the field vector.

The integro-differential equation (\ref{ke}) is equivalent to a system of three time-local ordinary differential equations
\begin{equation}
\label{ode}
    \dot{f}(\mathbf{p} ,t)  = \frac{1}{2}\lambda u(\mathbf{p} ,t) , \quad \dot{u}(\mathbf{p} ,t)  = \lambda (1-2f(\mathbf{p} ,t) ) - 2 \varepsilon v(\mathbf{p} ,t) , \quad \dot{v}(\mathbf{p} ,t) = 2 \varepsilon u(\mathbf{p} ,t)  ,
\end{equation}
where $u(\mathbf{p},t), v(\mathbf{p},t)$ are auxiliary functions describing vacuum polarization effects.

The distribution function $f(\mathbf{p} ,t)$ is zero in the in-vacuum state where the external field is absent. 
The KE (\ref{ke}) allows to describe the EPP evolution and to get the final distribution function 
$f_{\rm out}(\mathbf{p})$ in the out-state.
The duration of the action of the external field is limited, its time dependence is arbitrary.
The in- and out-vacuum states are different although $E_{\rm out}=E_{\rm in}=0$ due to the fact that $A_{\rm out}\neq A_{\rm in}$.

The total pair number density in the out-state is defined as
\begin{equation}
\label{dens}
    n_{\rm out} = 2 \int \frac{d\mathbf{p}}{(2 \pi)^3} f_{\rm out}(\mathbf{p}).
\end{equation}

\section{EPP creation in the few photon area}
\label{transit}

We will investigate the KE (\ref{ke}) numerically for the simplest Eckart-Sauter model of the one-sheeted electric field with characteristic duration  $T$ of its action defined by
\begin{equation} 
\label{field1}
E(t) = E_0 \cosh^{-2}(t/T),  ~~~\,  
A(t)= -{TE_0} \tanh(t/T) .
\end{equation}
A more realistic laser pulse model is the one with a Gaussian envelope \cite{8}
\begin{equation} 
\label{field2} 
E(t) = E_0  \cos{ (\omega t) }\ e^{-t^2/2\tau^2 }, ~~~ A(t) = -\sqrt{\frac{\pi}{8}} E_0\tau \exp{(-\sigma^2/2)}\;\text{erf}\left(\frac{t}{\sqrt{2}\tau} -i\frac{\sigma}{\sqrt{2}}\right) + c.c. , 
\end{equation}
where $\sigma= \omega \tau$ is a dimensionless measure for the characteristic duration of the pulse $\tau$ 
connected with the number of oscillations of the carrier field.

\begin{figure}[!h]
\includegraphics[width=0.48\textwidth]{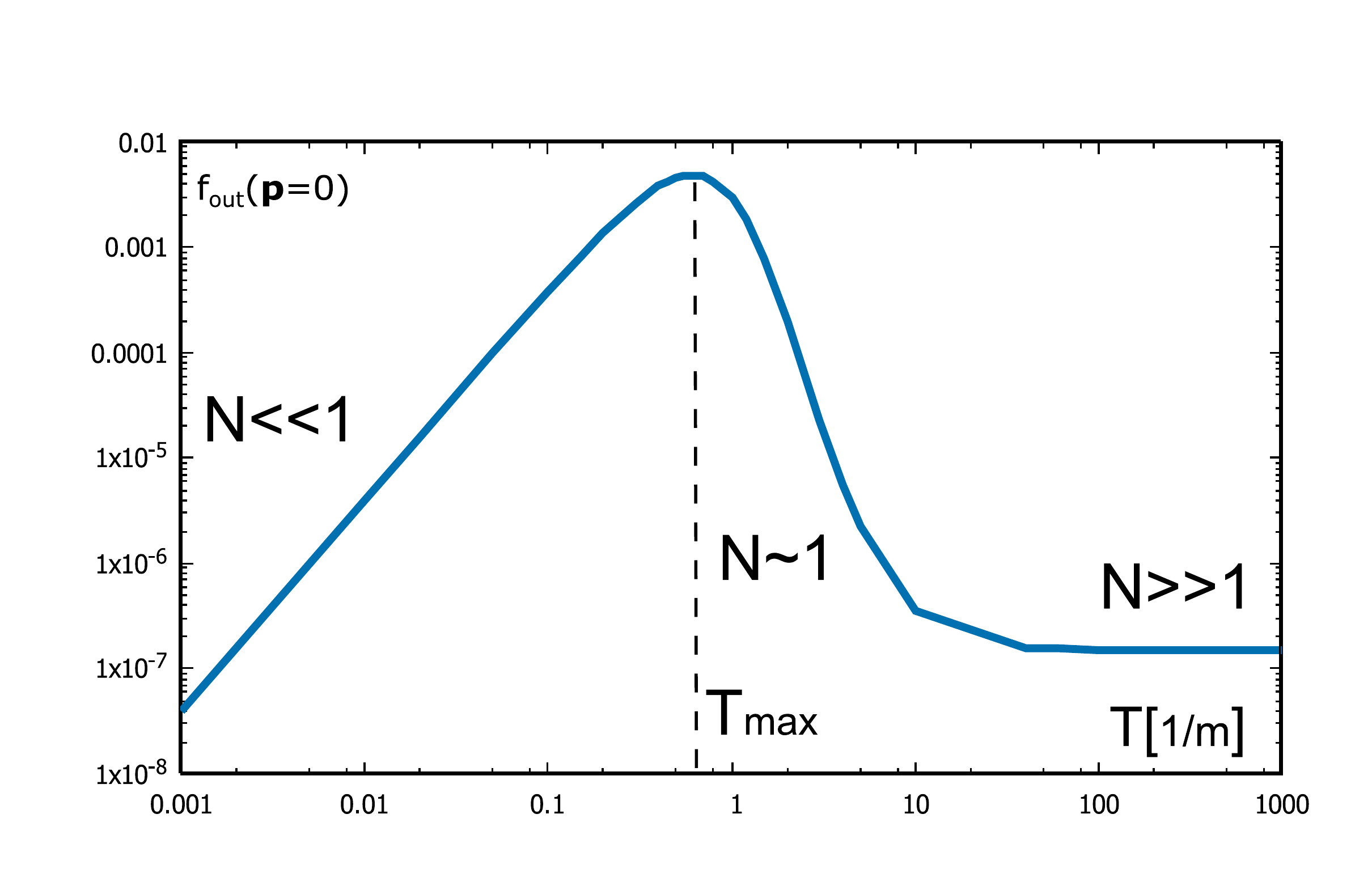} \hfill
\includegraphics[width=0.48\textwidth]{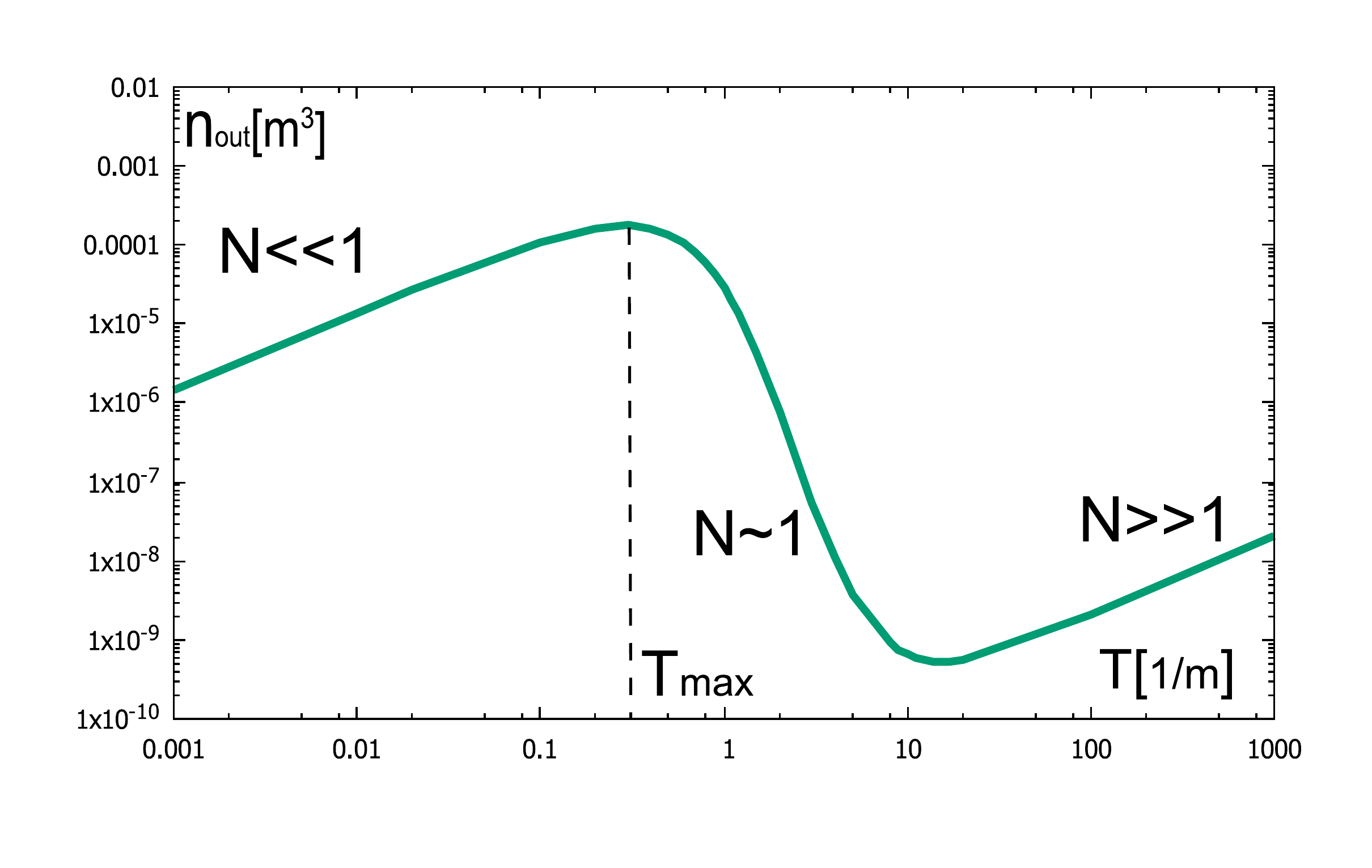}
\caption{Behavior $f_{\rm out}(\mathbf{p}=0)$ (Left panel) and density (\ref{dens}) (Right panel) in the tunneling area ($N \gg 1$), the few photon regime ($N \sim 1$) and the domain of high-energy photons 
($N \ll 1$) for the field model (\ref{field1}) with $E_0=0.2\, E_c$.
\label{fig:1}}
\end{figure}

In Fig.~\ref{fig:1} we demonstrate the resonant behaviour of the residual distribution function 
$f_{\rm out}(\mathbf{p})$ in the point $\mathbf{p}=0$ (left panel) and the total residual EPP number density $n_{\rm out}$ (\ref{dens}) (right panel) as functions of the field pulse duration $T$ of the field model (\ref{field1}).
The tendency of the function $f_{\rm out}$ to assume a constant value in the tunneling area $T \gg 1$ can be associated with the accumulation effect $n_{\rm out} \sim T$ (right panel).
This effect is well known from the exact solution of the problem \cite{9}.
A sharp decrease of the EPP creation efficiency is observed in the region of very small duration 
$T\, m\ll 1$. Analogous results can be obtained for the field model (\ref{field2}).

It would be natural to interpret these results in terms of the minimal photon number $N$ required to overcome 
the energy gap and to create an electron-positron pair.
The field models (\ref{field1}) and (\ref{field2}) themselves do not allow this because of the non-monochromaticity of these fields.
However, on the qualitative level one can connect the value $T_{max}\, m \simeq 0.5$ of the maximal efficiency of EPP creation with the Breit-Wheeler process of electron-positron pair creation by a pair of colliding photons.
Thus, $T_{max}\, m \sim N_{max} = 2$.
When extrapolating this relation to both sides, the large and small durations, we obtain $N = 4T\, m$ for the field model (\ref{field1}).
So, we can mark out on Fig.~\ref{fig:1} and next figures the border between few photon and tunneling areas.

\begin{figure}[!h]
\includegraphics[width=0.48\textwidth]{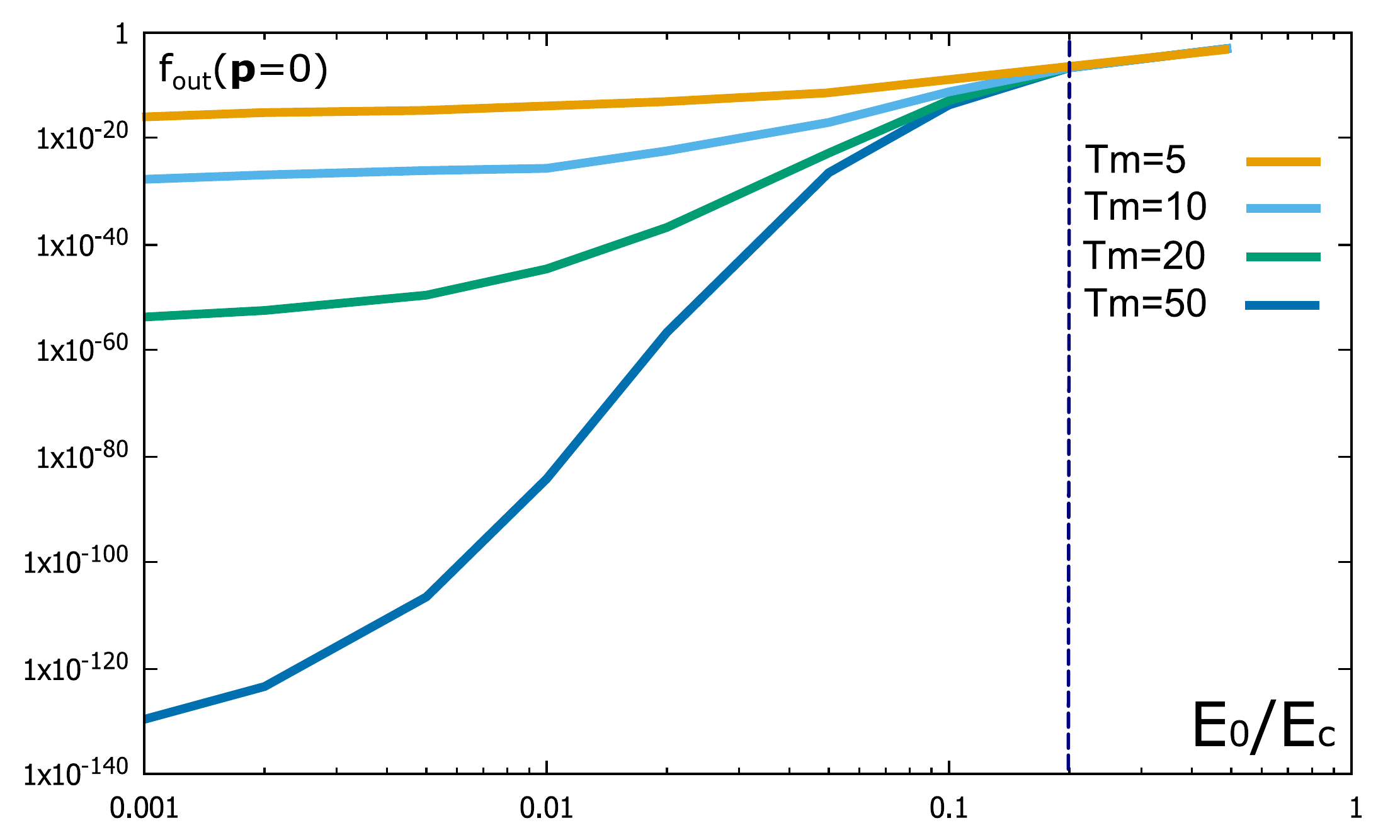}\hfill
\includegraphics[width=0.48\textwidth]{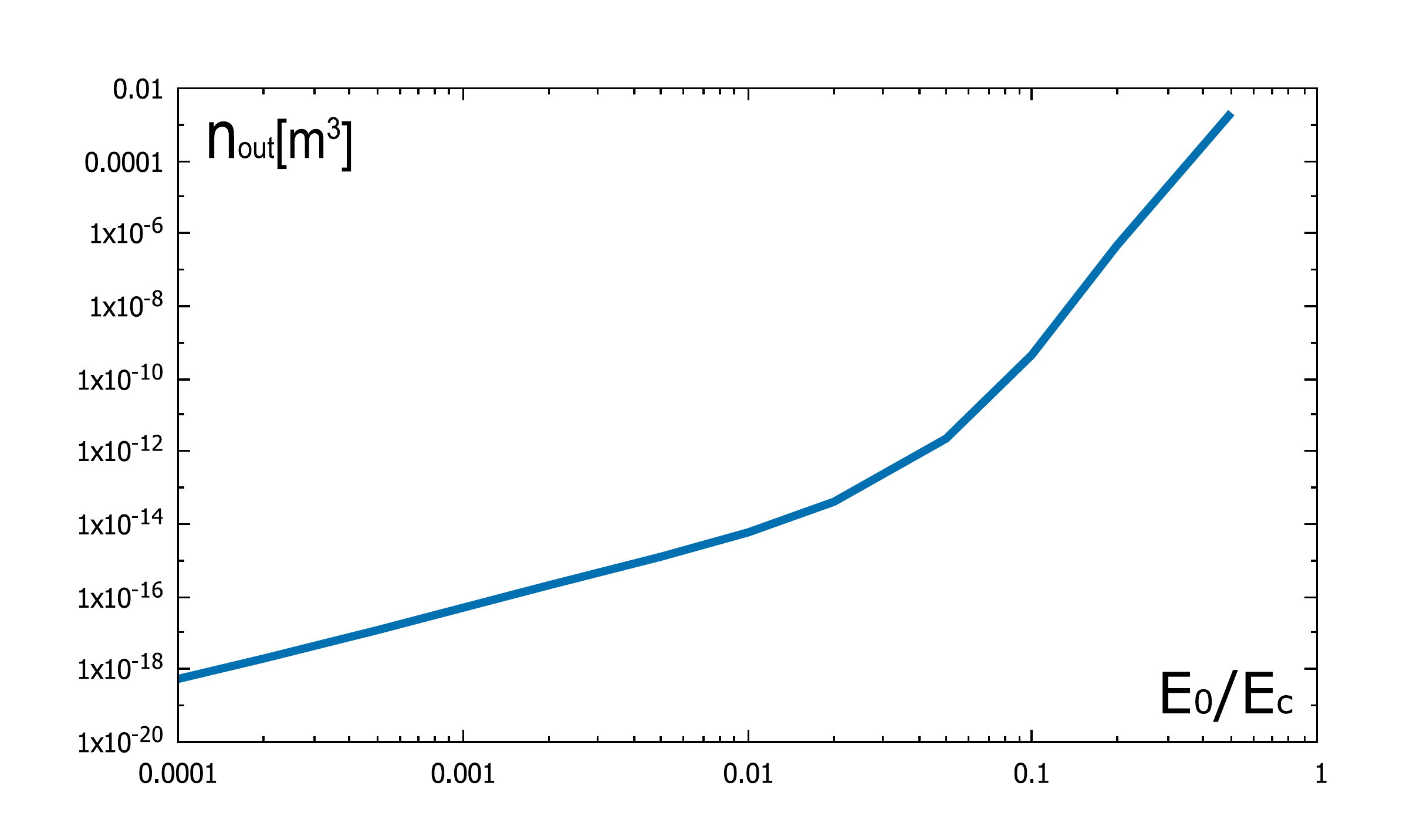}
\caption{
{\bf Left panel:} Relation between $f_{\rm out}(\mathbf{p}=0)$ and the amplitude of the external field $E_0$ for different values of pulse duration $T$.
{\bf Right panel:} The dependence of the density $n_{\rm out}$ on the external field $E_0$ in the few photon regime for $T\, m = 5$. 
\label{fig:2}}
\end{figure}

A similar correspondence can be introduced for the field model (\ref{field2}) in the case of a long pulse 
$\sigma \gg 1$, where $N = 2\, m/\omega$.

The existence  of the maximum in the two photon domain correlates with the result of standard perturbation  theory  for $E_0/E_c \ll 1$ in the few photon region where this process is dominant \cite{11}.

The field strength dependence of the efficiency of EPP creation is depicted in Fig.~\ref{fig:2}. 
The left panel shows that sizeable values of the distribution function $f_{\rm out}(\mathbf{p}=0)$ can be reached for moderate field strengths when entering the few photon domain.

\begin{figure}[!h]
\includegraphics[width=0.48\textwidth]{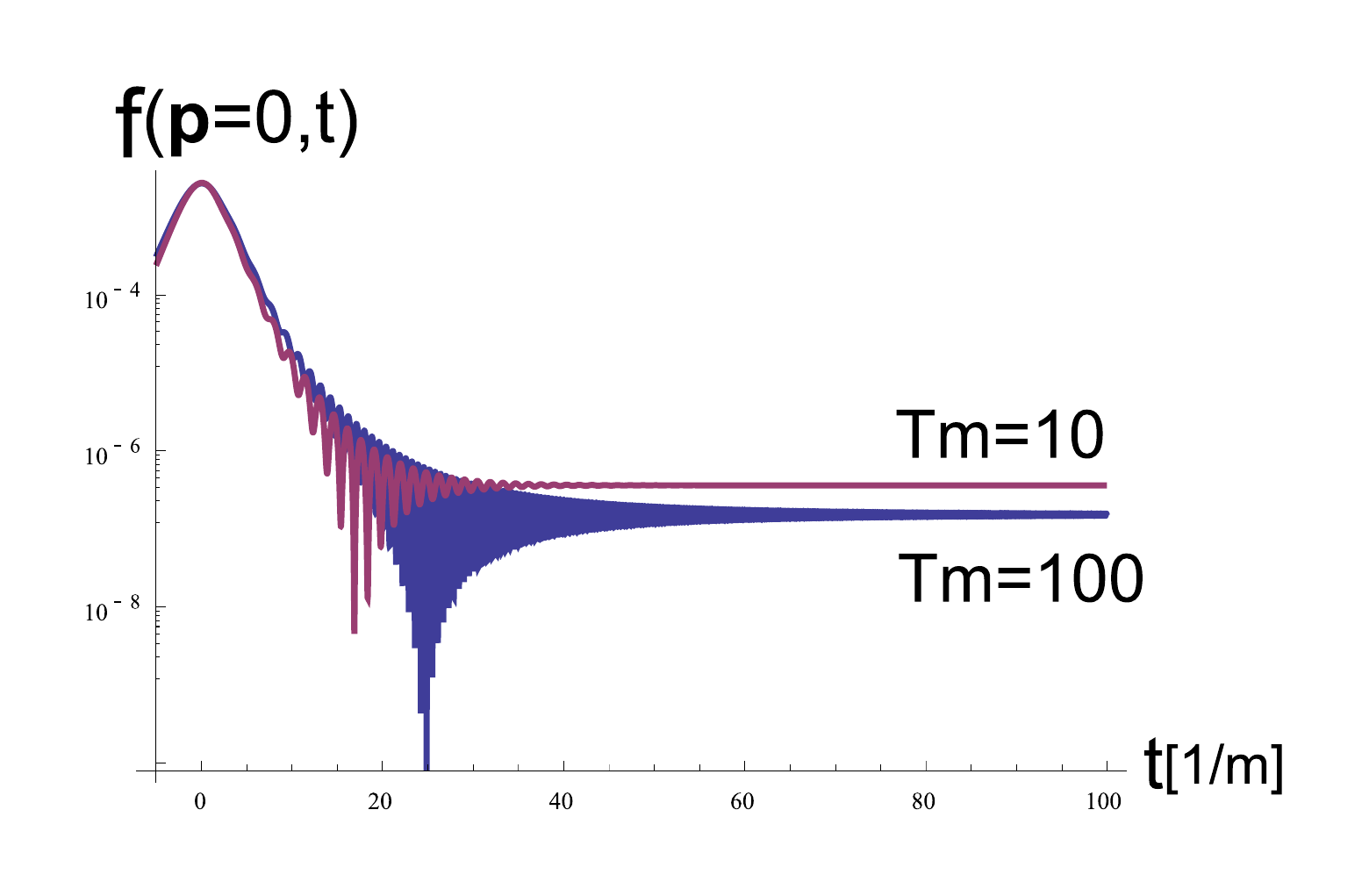} \hfill
\includegraphics[width=0.48\textwidth]{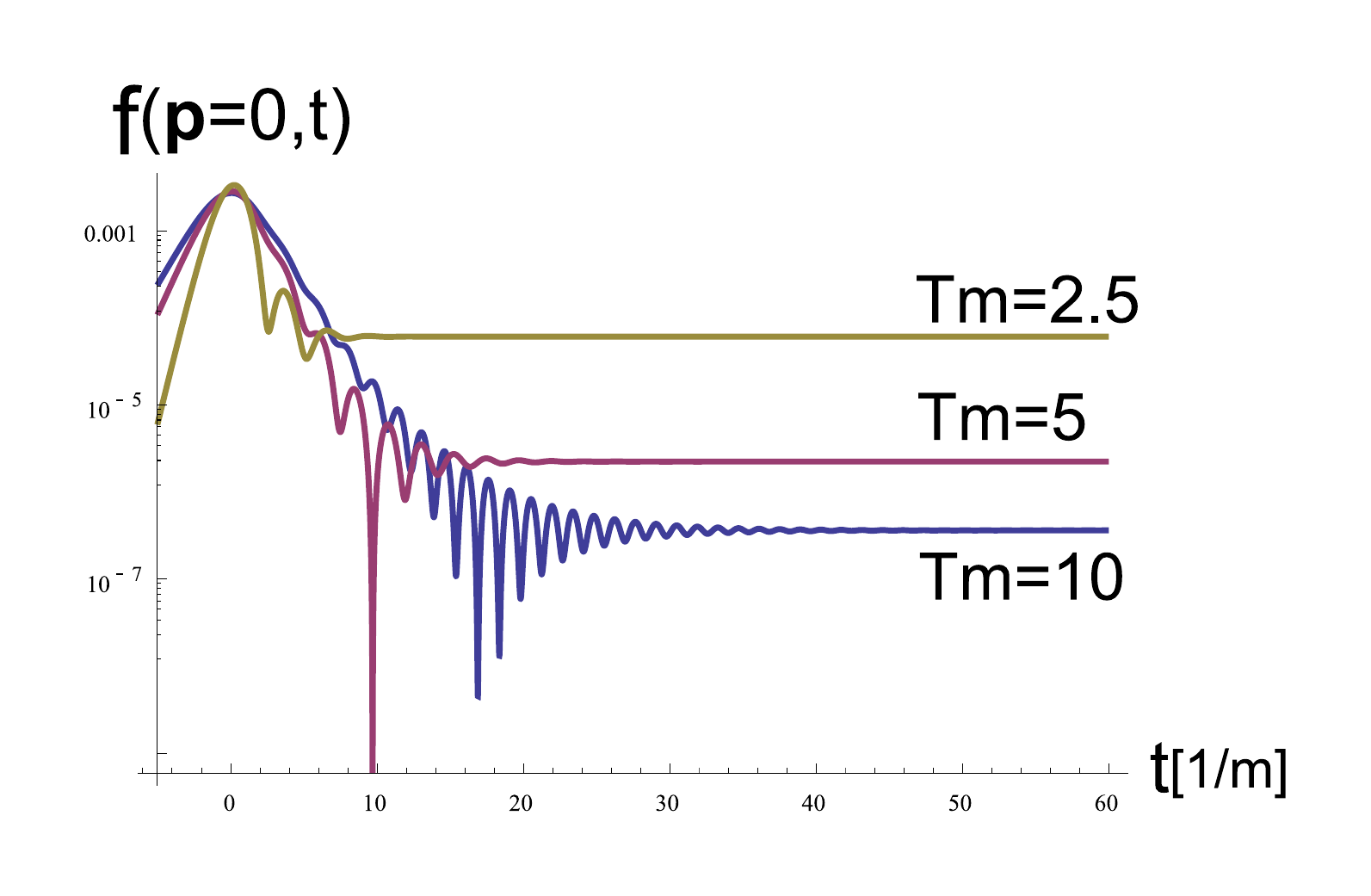}
\caption{The transition from the quasiparticle EPP to the final out-state for the Eckart-Sauter pulse (\ref{field1}) 
with $E_0 = 0.2\, E_c$ at different values of $T$ and $\mathbf{p}=0$. 
{\bf Left panel:} Transfer from the tunneling ($T\, m=100$) to the few photon area ($T\, m=10$).
{\bf Right panel:} Few photon area at different $T$. 
\label{fig:3}}
\end{figure}
The right panel of Fig.~\ref{fig:2} shows the dependence of the created EPP density on the field strength $E_0$ in the few photon regime at $T\ m=5$.
In the immediate neighborhood of the critical field $0.05\, E_c \lesssim E_0 \lesssim 0.5\, E_c$ 
the efficiency of the pair creation process decreases very quickly (exponentially, as in the tunneling area). But at $E_0 \lesssim 0.02\, E_c$ and smaller it  decreases as $E_0^2$.
This promises an observable density of EPP even at field strengths several orders of magnitude smaller than $E_c$.

The left panel of the Fig.~\ref{fig:3} shows a behavior of the distribution function that is typical for the phase transition from the vacuum to a final state of the EPP in the tunneling region ($T\, m=100$) for a one-sheeted external electric field (\ref{field1}).

These are the three stages of evolution \cite{1}: 
(i) quasiparticle EPP in the region of the maximal values of the external field, 
(ii) the transition region with rapid oscillations of the distribution function and 
(iii) the final state approaching a constant residual value $f_{\rm out}$.

Moreover, in the tunneling domain the saturation of the distribution function for a fixed point in momentum space appears.
For example, the value of $f_{\rm out}$ at a given point $\mathbf{p}=0$ does not depend on $T$ and is defined by the field amplitude $E_0$ only.
The left panel of Fig.~\ref{fig:3} shows the deviation from this rule for $T\ m=10$.
The right panel shows a further sharp increase of $f_{\rm out}$ at the same $E_0$ only due to the lowering of $T$.
On observes a gradual narrowing and a disappearance of the fluctuations in the transient domain.
This is due to the fact that the role of the vacuum polarization effects decreases with decreasing $T$.

\begin{figure}[!h]
\includegraphics[width=0.48\textwidth]{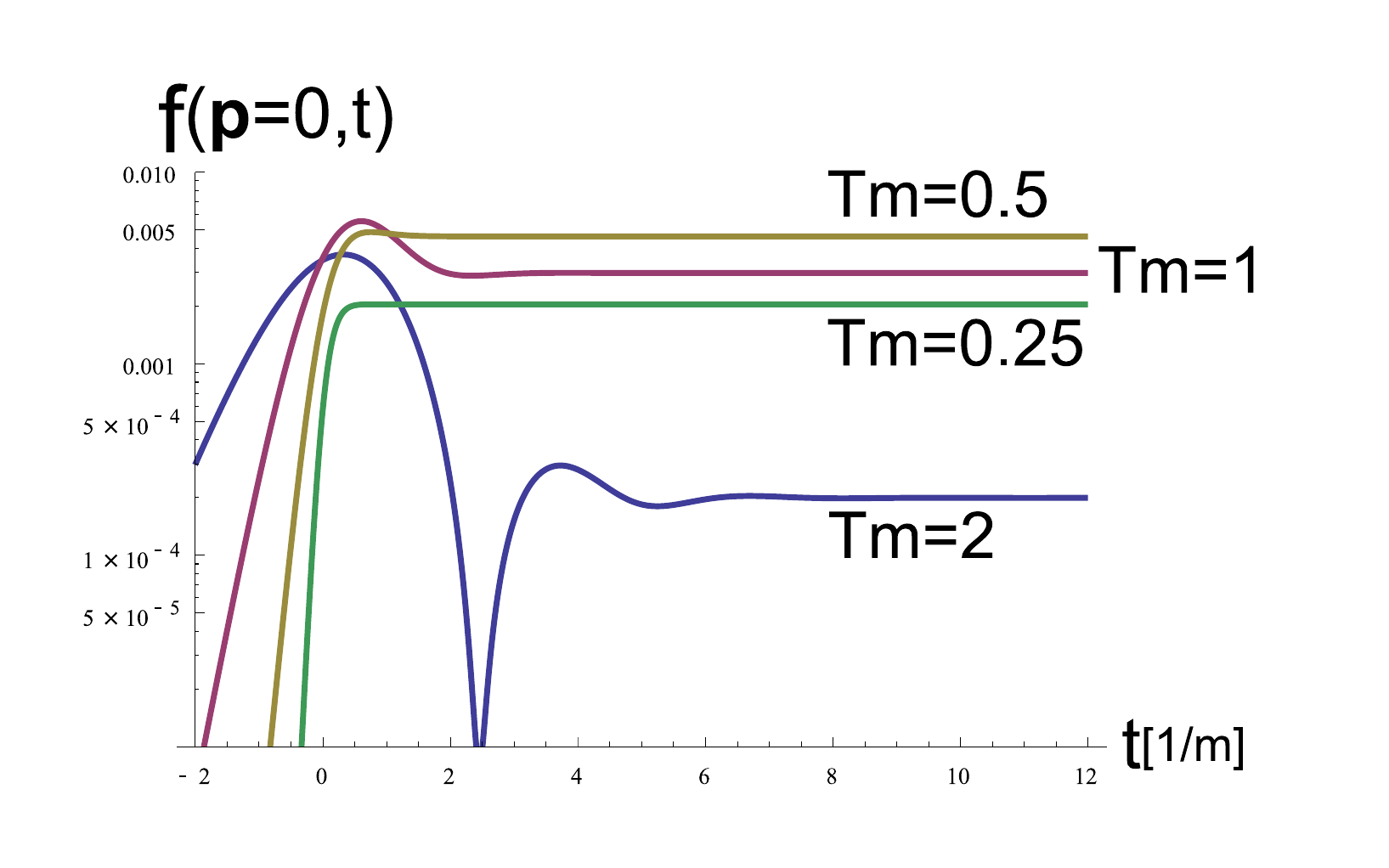} \hfill
\includegraphics[width=0.48\textwidth]{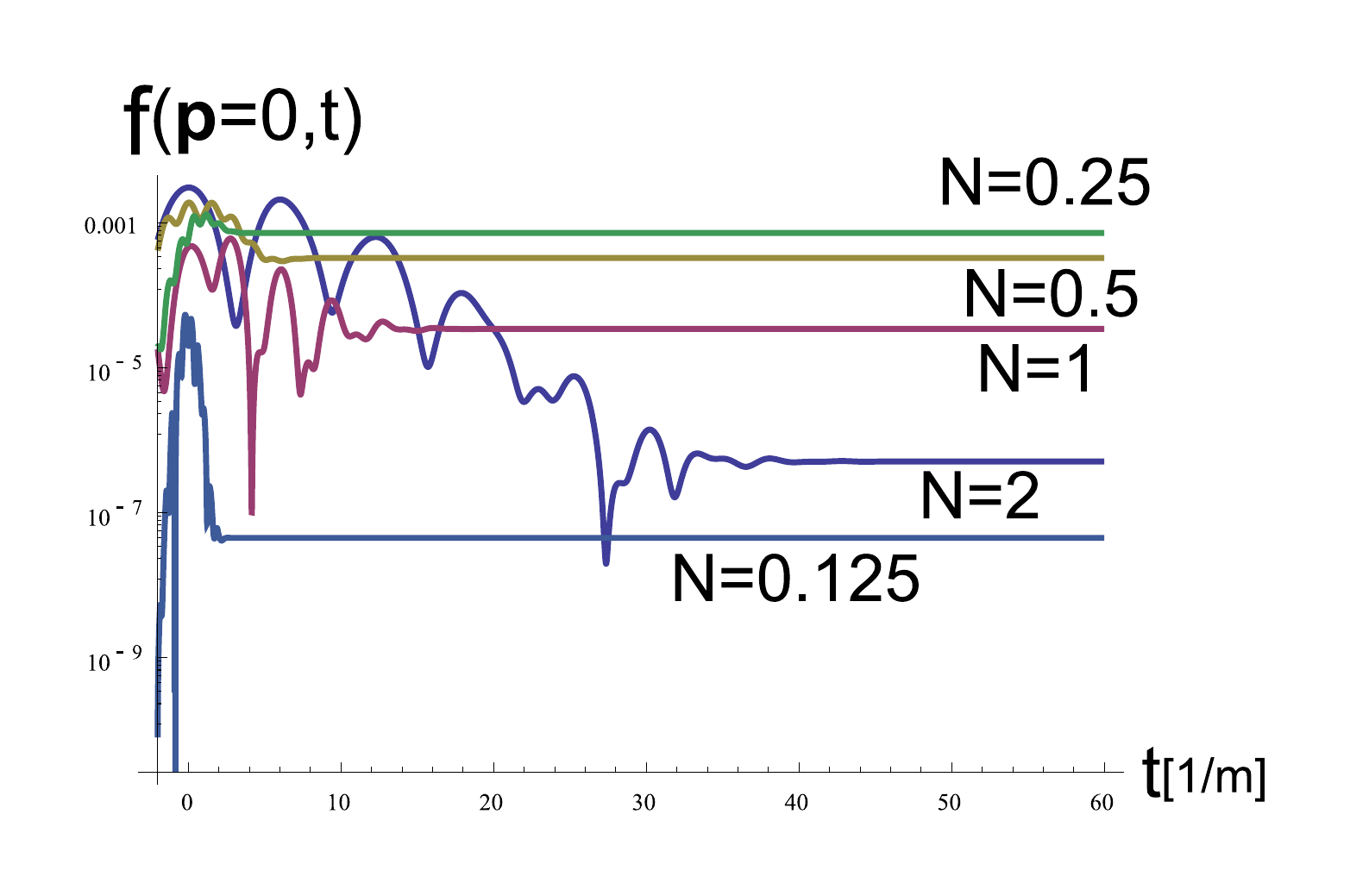}
\caption{At a fixed electric field $E_0$ the maximum of $f_{\rm out}$ ($\mathbf{p}=0$) occurs in the vicinity of $T(N) \sim 1/m$. 
Entering further into the region with high photon energies results in a reduction of the efficiency of the pair creation  process.
{\bf Left panel:} Features of the evolution of the distribution function in the vicinity of $T\sim 1/m$ for the field (\ref{field1}).
{\bf Right panel:} Similar results for field (\ref{field2}) in the vicinity $N \sim 1$ and $\sigma = 5$. 
\label{fig:4}}
\end{figure}

Fig.~\ref{fig:4} shows that both field models (\ref{field1}) and (\ref{field2}) exhibit a rapid growth of 
$f_{\rm out}$ in the few photon domain.
However, near $T(N) \sim 1/m$ the growth stops and turns to a decrease for further increasing  energy of the photons of the external field.
There is a complication of the character of the evolution of the distribution function in the case of a more complex time dependence of the field (\ref{field2}) (right panel of Fig.~\ref{fig:4}).
Here one observes also a gradual attenuation of oscillations in the transient region.

\section{Conclusion}
\label{discuss}

The results obtained in this contribution show that in the transition from the regime of tunneling (constant field and a field with a relatively slow time-dependence) to the few photon regime (fastly alternating fields with a characteristic frequency comparable to the Compton frequency) there are qualitative changes in the characteristics of the process of EPP creation from vacuum under the action of an external electric field. 
The evolution of the distribution function from the initial vacuum state to the final state loses a specific transient region of strong fluctuations. 
The transition to the few photon region is accompanied also by a considerable growth of the pair production effectiveness.
So for the field strength $E_0=0.2\, E_c$ the pair density in the out state increases $3.34\times10^5$ times.
But more importantly, the effect of pair production in the few photon domain can still be detected for much weaker fields, with field stengths several orders of magnitude smaller than $E_c$.

In contrast to the works on the bifrequent mechanism of vacuum EPP excitation \cite{2,3} the present investigation has shown that the transition in the few photon area ensures in itself a high intensity of the EPP creation process without participation of the tunneling mechanism.
This is the main result of this work.
It would be highly topical to perform a detailed comparison of the results of these two approaches.

Furthermore, it would be interesting to compare the results of the kinetic approach with the approach based on direct calculations of the probability of the Breit-Wheeler process in the framework of the S-matrix formalism \cite{10}.

\subsection*{Acknowledgements}
S.A.S. and A.D.P. are grateful to the organizers of the XXIII. Baldin Seminar for the support of their participation at this conference. The work of S.A.S., D.B.B. and L.J. was supported in part by the Polish National Science Centre (NCN) under grant No. UMO-2014/15/B/ST2/03752. 
D.B.B. was supported by the MEPhI Academic Excellence Project under contract No. 02.a03.21.0005.
L.J. acknowledges support from the Bogoliubov-Infeld program for his visits to JINR Dubna in 2016.

\end{document}